\documentclass{article}

\usepackage{arxiv}
\usepackage{natbib}
\setcitestyle{square,sort,comma,numbers}
\usepackage[utf8]{inputenc} 
\usepackage[T1]{fontenc}    
\usepackage{hyperref}       
\usepackage{url}            
\usepackage{booktabs}       
\usepackage{amsfonts}       
\usepackage{nicefrac}       
\usepackage{microtype}      
\usepackage{lipsum}
\usepackage{framed,multirow}
\usepackage{booktabs}
\usepackage{algorithm}
\usepackage{algorithmic}%
\usepackage{bbm}
\usepackage{lineno}
\usepackage{amssymb}
\usepackage{latexsym}
\usepackage{amsmath}
\usepackage{subfigure}
\usepackage{placeins}
\usepackage{hyperref}
\usepackage{float}
\usepackage{pdfcomment}
\usepackage{graphics}
\usepackage{graphicx}
\usepackage{xcolor}
\definecolor{newcolor}{rgb}{.8,.349,.1}
\definecolor{newcolor}{rgb}{.8,.349,.1}
\definecolor{mygreen}{rgb}{0.1, 0.75, 0.15}

\newcommand{\tabitem}{~~\llap{\textbullet}~~}

\newcommand{\removeGG}[1]{}

\newcommand{\removeRH}[1]{}

\title{Feature Extraction for Temporal Signal Recognition: An Overview}

\author{
  Imad Rida  \\
  Normandie Univ, UNIROUEN, UNIHAVRE, INSA Rouen, LITIS \\
  76 000, Rouen\\
  France \\
  \texttt{imad.rida@insa-rouen.fr} \\
}

\begin{document}
\maketitle

\begin{abstract}
Due to the huge progress of the recording devices, data from heterogeneous nature can be recorded, such as spatial, temporal and spatio-temporal. Nowadays, time-based data is of particular interest since it has the ability to capture the characteristics evolution of the data over time. The temporal data could be gait, auditory scene, piece of music, and so on. In this paper, we are particularly interested in feature extraction for two different temporal recognition applications namely, audio and human behavior analysis and recognition. Indeed, relevant and discriminative features are of critical and fundamental importance to achieve high performances in any automatic pattern recognition system. This work is intended to provide researchers with a brief overview of the different existing features through an understanding of basic taxonomies which may serve as a reference to identify the adequate features for a specific task.
\end{abstract}


\section{Introduction}

Over the past two decades, there has been a massive and abundant amount of data garnered from social media, data from internet-enabled devices (including smartphones and tablets), video and voice recordings (digital cameras, microphones), etc. The recorded data represents a huge and important resource of information and knowledge which could be exploited in real life applications such as, security, education, healthcare etc \cite{al2018palmprint}. 

Despite the ability of recorded data to give useful information, it is not always captured in ready and adequate format for analysis and interpretation which clearly shows the need of novel efficient methods to address this problem \cite{rida2018palmprint,shariatmadari2018off}. However, doing this correctly and completely represents a continuous challenging problem which took the effort and attention of researchers. Temporal signals constitute a popular class of signals, where data records are indexed by time. There is a large variety of examples in the context of  temporal signal recognition applications; within the most popular ones we can find: audio signal recognition or human behavior analysis and recognition.

Section \ref{1} introduces audio signal recognition.  Section \ref{2} presents human behavior analysis and recognition. Section \ref{3} describes the architecture of an automated recognition system. Section \ref{4} explains audio feature extraction. Section \ref{5} reports human behavior analysis and recognition features extraction. Finally, Section \ref{6} concludes our paper.



\section {Audio Signal Recognition}
\label{1}
Human listeners are very good at all kinds of sound detection and identification tasks, from understanding heavily accented speech to noticing a ringing phone underneath music playing at full blast. Efforts to duplicate these abilities on computer have been particularly intense in the area of audio signal recognition. The beginning was with speech-based applications \cite{speech2}, later extended to other audio recognition tasks, ranging from music analysis \cite{muller2011signal} to the problems of analyzing the general "ambient" audio \cite{rossi2013ambientsense}.

\begin{figure}[!ht]
\centering
\includegraphics [width= 9 cm] {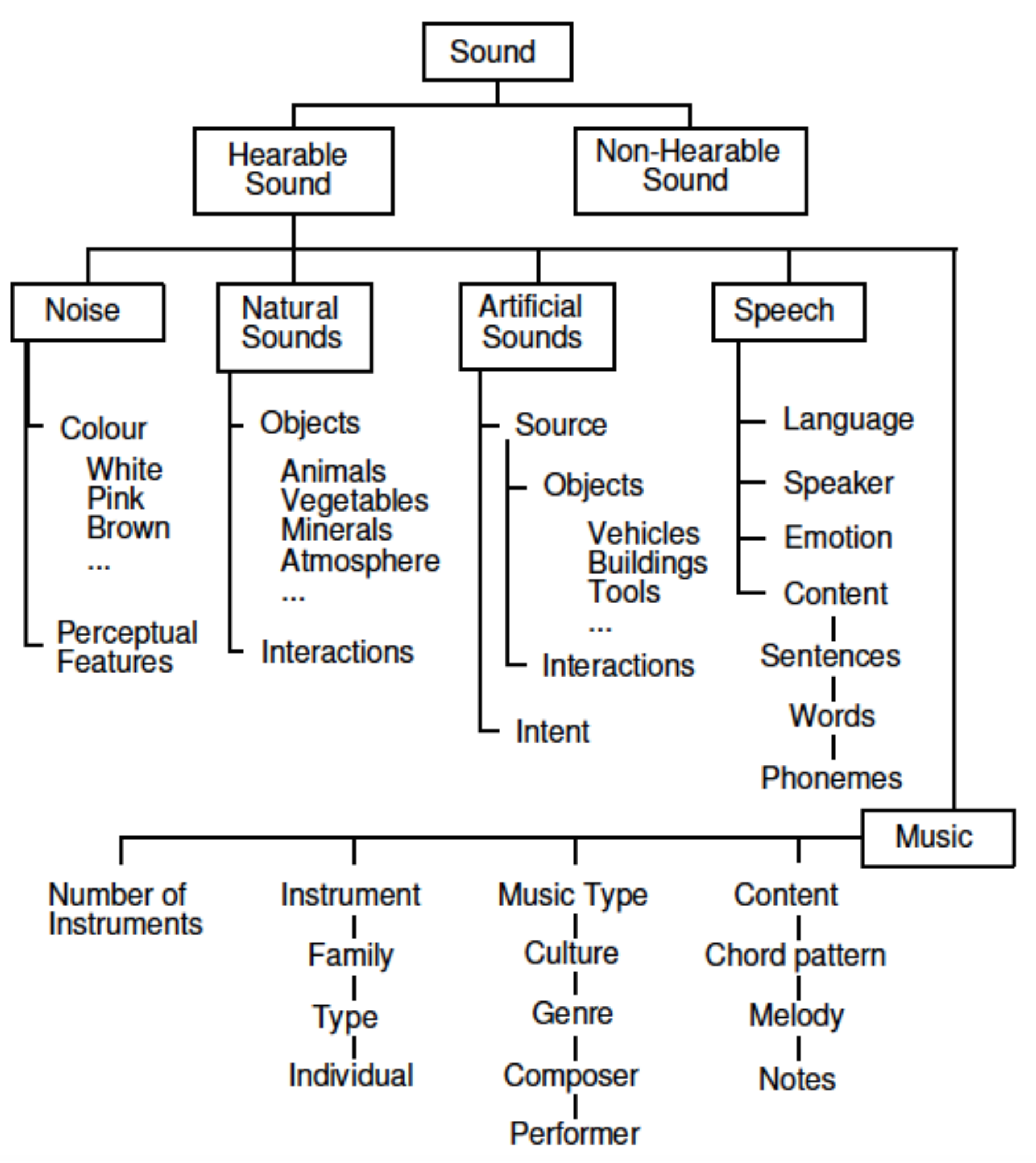}
\caption{Taxonomy of sounds \cite {gerhard}.}
\label{fig:taxanomy}
\end {figure}

To tackle the problem of audio signal recognition, a development of auditory signals taxonomy is needed. Gerhard \cite {gerhard} defines the sound as a pattern of air pressure that is detectable (the average human can hear frequencies between 20 Hz and 15 000 Hz). He splitted  the hearable sound into 5 main categories: noise, natural sounds, artificial sounds, speech and music as is shown in Figure \ref {fig:taxanomy}.

From this taxonomy one can derive a broad range of audio signal recognition problems based on speech, music and mixture of artificial and natural sounds. In the following we briefly emphasize on three prominent applications namely speech recognition, music transcription and computational auditory scene analysis. These trends of research aim at building intelligent machines able to interpret and infer based on audio information.

 \subsection{Speech} 
 Speech has been one of the fundamental audio research topics for many years now. There are three main topics in speech research in recognition context: speaker, speech and language recognition. Speaker recognition is the general term of discriminating one person from another based on the sound of their voices. It was for instance a good biometric modality used as  alternative of conventional passwords, personal identification numbers (PINs) or smart cards \cite{speaker1,speaker2,speaker3}.  Speech recognition is the ability of a machine to convert a speech signal to a readable sequence of words and phrases \cite {speech1,speech2,xiong2016achieving}, while language recognition refers to the process of automatically identifying the language spoken in a speech sample \cite {language1,language2}.
 
 \subsection{Automatic Music Transcription} 
In the past years, the problem of Automatic Music Transcription (AMT) has known an increased interest due to many applications associated with it, such as, interactive music systems, automatic search and annotation of musical information, as well as musicological analysis \cite{transomportance1,transomportance2}. It corresponds to the process of taking a sequence of sound waveform and extracting from it some form of musical notation related to the high-level musical structures \cite {amt}.  AMT machine generally follows three main stages, spectral estimation, pitch detection and symbol formation \cite {gerhard}.  Spectral estimation is usually done with Fourier analysis and the detected pitch information is represented in  recognizable format by humans and computers such as Music Instrument Digital Interface (MIDI). A melody line represented by a series of pitches could be represented in any key signature.

The AMT problem can be divided into several subtasks such as, musical instrument identification which seeks to identify the musical instrument(s) playing in a music piece \cite {instrument1,instrument2}; onset detection which aims to find beginnings of notes or events \cite{onset1,onset2} or music chord recognition \cite {chroma,chord,oudretemplate,rida2014supervised}. The latter represents the most fundamental structure and back-bone of the tonal system which makes them deft to represent occidental music. Moreover harmonic informations extracted from chord recognition task can serve as features for high level tasks such as music genre classification or music retrieval.

\subsection{Computational Auditory Scene Analysis}  
Perception refers to the process of becoming aware of the elements of the environment through physical sensation, which can include sensory input from the eyes, ears, nose, tongue, or skin.  While most of the efforts have focused on vision perception (it represents the dominant sense in humans to build intelligent artificial machines), there is now a growing interest based on audio modality. Computational Auditory Scene Analysis (CASA) refers to the computational analysis of an acoustic environment, and the recognition of specific sounds and events in it. Automatic sound event detection (also called acoustic event detection) and Computational Audio Scene Recognition (CASR) represent two emerging topics in the general context of CASA \cite {casa}. The former aims to process the continuous acoustic signals and convert them into symbolic descriptions of the corresponding sound events present at the auditory scene when the latter seeks to recognize the acoustic environment or context. Applications that can specifically benefit from CASA include automatic tagging in audio indexing \cite{mesaros2010}, context-aware services \cite {aware}, intelligent wearable devices \cite{wearable} and robotics navigation systems \cite{robotics}.

\section{Human Behavior Analysis and Recognition}
\label{2}

There is an increasing interest in video surveillance applications to propose solutions able to analyze the human behaviors and identify individuals. Currently, visual surveillance is one of the most active research areas in computer vision and pattern recognition. The goal of visual surveillance is not only to replace the human eyes by cameras but also to make the surveillance task as automatic as possible. Applications in visual surveillance can be divided into two main tasks, human behavior analysis and person recognition.

\subsection{Human Behavior Analysis}
In the past years, a considerable number of surveillance cameras have been installed in public places, train stations, airports and many research efforts have been devoted to build intelligent systems able to analyze the visual data in order to extract information about the humans behavior in scenes. Ideal intelligent monitoring system should be able to automatically, analyze the collected video data, detect the suspicious or endangering behaviors and give out an early warning before the adverse event happens.

\begin{figure}[!ht]
\centering
\includegraphics [width= 9 cm] {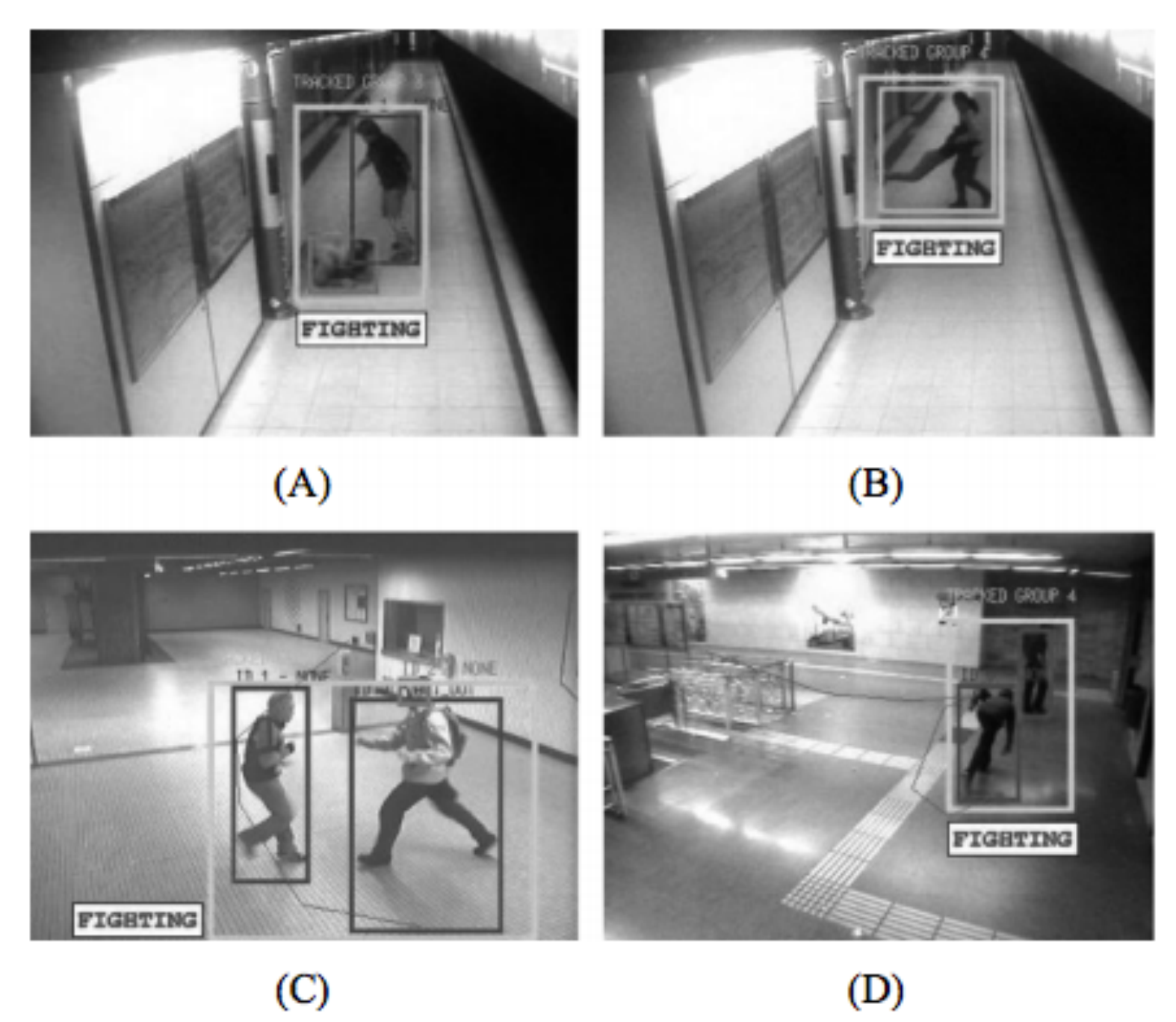}
\caption{Example of suspicious (fighting) behave detection \cite {behave4}.}
\label{fig:susp1}
\end {figure}

\begin{figure}[!ht]
\centering
\includegraphics [width= 9 cm] {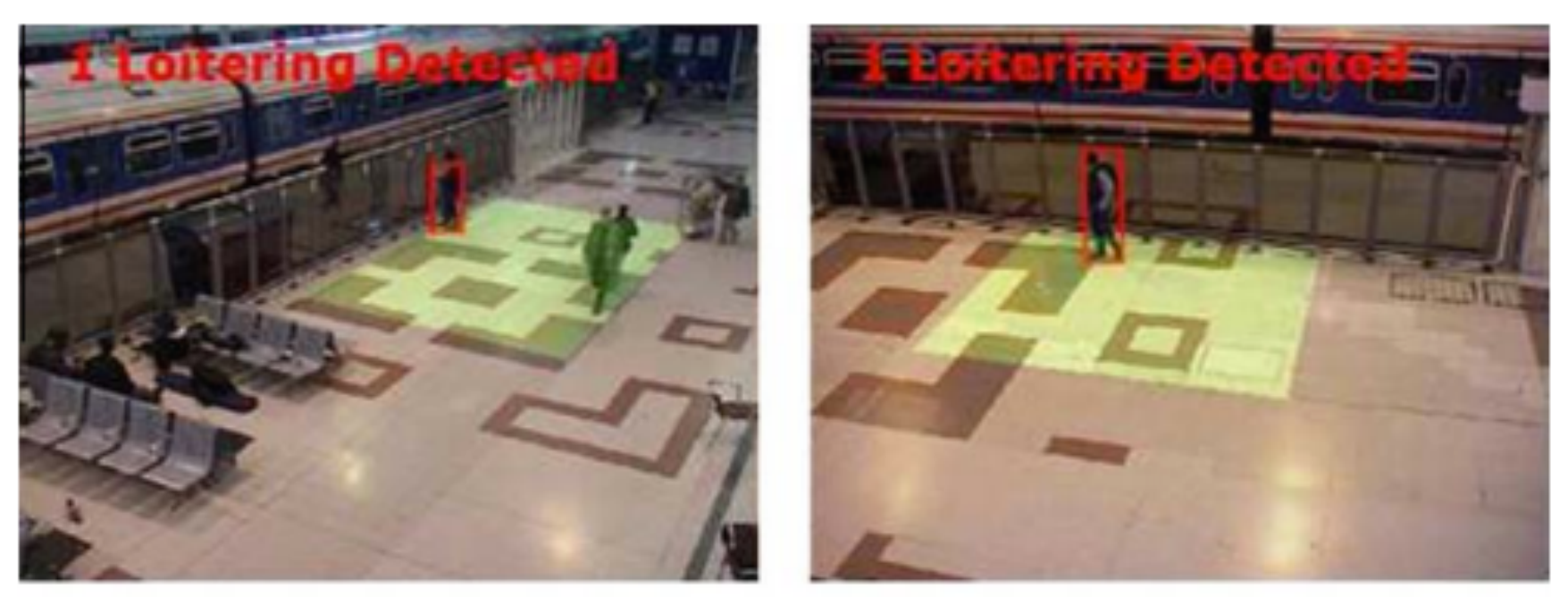}
\caption{Example of suspicious (loitering) behave detection \cite{lim2014}.}
\label{fig:susp2}
\end {figure}

Many suspicious behaviors could be defined depending on the application domain, such as loitering (waiting time to catch a bus longer than a threshold time) illustrated in Figure \ref {fig:susp1} or fighting shown in Figure \ref {fig:susp2}. Detection of suspicious human behavior involves modeling and classification of human activities based on predefined knowledge. However this task is not trivial due  to the randomness and complex nature of human movement \cite {behave3,behave4,behave1,behave2}.

\subsection{Human Recognition in Surveillance Systems}

A system which detects abnormal behavior should also be able to identify all the suspicious persons in the scene, and track them across the zones.  Monitoring system requires not only to estimate the location and behavior, but also to obtain the identity information \cite{rida2015improved}. Gait is the most suitable biometric modality in the case of intelligent video surveillance \cite{hayfron2003,rida2019robust}. In monitoring scenes, people are usually distant from cameras, which makes most of biometric features not suitable even the use of face for identification. The drawbacks are obvious, for example, view angle variations and occlusions cause the impossibility to capture the full faces and distance brings low-resolution face images. Therefore, face can not always achieve good performances in practice. In contrast, gait is a behavioral biometric, including not only individual appearance, such as limb, leg length, width, but also the dynamic information of individual walking \cite{rida2016human,rida2015unsupervised}. Compared with other biometric modalities, gait is remote accessed and difficult to imitate or camouflage. Moreover, the capturing process does not require cooperation, contact with special sensor, or high images resolution  \cite {gait1,gait2,rida2016gait}.


\section {Architecture of Automated Recognition Systems}
\label{3}

Assume that we have several objects associated with classes and that objects belonging to the same class share the same features more than with objects in other classes. The pattern recognition problem consists of assigning a new unlabeled object to a class.  It is accomplished by determining the features of the object and identifying the class of which those features are most correlated.

Given the goal of recognizing objects based on their features, the main task of an automated recognition system can be divided into three basic subtasks \cite{rida2018palmprintt}: the description subtask which generates features of an object using feature extraction techniques, mapping raw features into another discriminative space where objects from different groups are well separated by feature representation techniques and finally the classification subtask which assigns a class label to the object based on those features and a trained classifier \cite{rida2018ensemble}.

As the ultimate goal of an automated recognition system is to discriminate the class membership of the observed novel objects, a good functional automated pattern recognition system should be able to classify the novel observed objects with the minimum misclassification rate possible. Relevant and discriminative features are of critical and fundamental importance to achieve high performances in any automatic pattern recognition system \cite{fei2017enhanced}.  Feature extraction seeks to transform and fix the dimensionality of an initial input raw data to generate a new set of features containing meaningful information contributing to assign the observations to the correct corresponding either on training samples or new unseen data class. 

Different type of information can be extracted from the initial recorded raw data (time, frequency, spatial information etc) depending on the nature of the input raw data, the context and domain of the task. In the following we present a general overview of the commonly used features in the domain of audio and human behavior analysis application.

\section{Audio Features Extraction}
\label{4}

Humans have  powerful brain capabilities to analyze and distinguish between different sounds and assign them to a specific semantic class. Unfortunately this is not possible for the machines due to the hidden nature of semantic information in the recorded sounds. This motivates the researchers to introduce several processing tools for audio signal which led to a large variety of features for different applications, such as music transcription, CASA, speech recognition etc. 

Feature extraction is of extreme importance since the performance of the system depends on the quality of the extracted features. The features, determine which information and properties are available during the recognition process. They should capture enough invariant audio properties within the same class and variant ones between different classes. 

Audio features represent specific characteristics of audio signals. Several attributes have been introduced to describe different types of audio signals from psychoacoustic point of view such as, duration, loudness, pitch, and timbre \cite {mitrovic}. 

\begin{description}
\item Duration: represents the time between the beginning and the end of the audio signal. The envelope of the sound over time can be divided into, Attack, Decay, Sustain and Release (ADSR). 
\item Loudness: is a psychoacoustic property of the sound, it represents our human perception of how loud or soft sounds of various intensities are. The loudness of a sound is subjective, it varies from person to person and measured by sone and phon units \cite{sonephon}. 
\item Pitch: is a perceptual property. In \cite {pitch1} is defined as the intensive attribute of auditory sensation in terms of which a sound may be ordered on a scale extending from soft to loud. The pitch is measured with mel unit. In some cases the pitch means the fundamental frequency \cite{pitch2}.                
\item Timbre: is defined as the attribute of auditory sensation which makes the listener able to judge that two non-identical sounds which are presented similarly and have the same loudness and pitch are dissimilar \cite{pitch1}. It is the most complex attribute in the sound. For example, timbre helps to distinguish between two different instruments playing the same note with same loudness. 
\end{description}

Audio features extraction attempts to capture the aforementioned attributes most adapted to the application domain.  Audio features hold five main properties \cite {mitrovic}: signal format, domain, temporal scale, semantic meaning, and the underlying model which will be further discussed in the following.

\begin{itemize}
\item Signal format: there are two main categories, features based on linear coding and based on lossy compression. The majority of audio features are linearly coded based, however several works tried to introduce features in lossy compression context (MPEG format) \cite {mpeg}.

\item Domain: it represents the final domain of the extracted audio feature. The features could belong to different domains such as, temporal, frequency, cepstral, modulation frequency and reconstructed phase space \cite {mitrovic}. 

\item Temporal scale: in this property, the features could belong to three different categories, intraframe, interframe and global. In the intraframe features, the signal is considered locally stationary. Each frame is taken in consideration separately which results in one feature vector by frame. A well known example of intraframe (or short-time) features is MFCCs. In contrast the interframe features capture the temporal change of a given audio signal. An example of the interframe features are rhythmic features. Note also the global features which are computed from the whole signal.
                
\item Semantic meaning: it includes perceptual features which are based on the aspects of human perception such as pitch, rhythm, and physical features describing the audio signals based on physical and statistical properties (Fourier transform). 

\item Underlying model: there are two types of features, those based on psychoacoustic model and those without it. An example of psychoacoustic model is the incorporation of the filter banks \cite {mitrovic}.
\end{itemize}

\begin{figure*}[!ht]
\centering
\includegraphics [width= 12 cm] {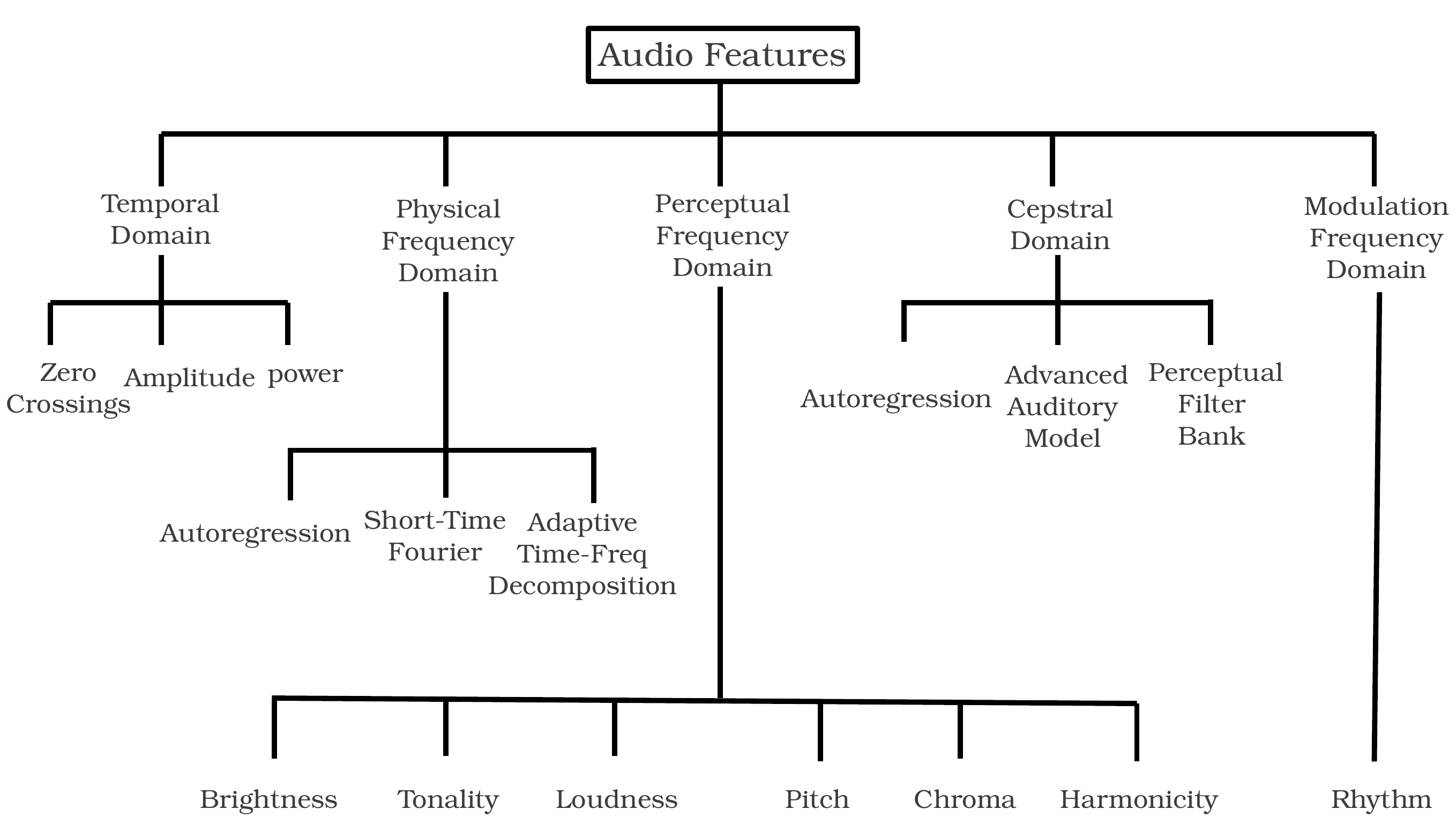}
\caption{Taxonomy of audio features.}
\label{fig:taxanomyy}
\end {figure*}

From the previous description one can remark there is a various and large variety of features to tackle the problem of audio signal recognition. This shows the need to a taxonomy organization into hierarchical groups with shared properties. Inspired by the taxonomy proposed by \cite{mitrovic}, we introduce the following organization which divides the audio features into five main domains, temporal, physical frequency, perceptual frequency, cepstral and modulation frequency as illustrated in Figure  \ref {fig:taxanomyy}. \\

\subsection{Temporal Features}
Temporal features are directly extracted from the audio raw data without any transformation. The temporal features include:
\begin{itemize}
\item Zero crossings: it is a very simple characteristic of the audio signals that has been used in speech recognition. We can find features as, Zero Crossing Rate (ZCR) \cite {zcr}, Linear Prediction Zero Crossing Ratio (LP-ZCR) \cite{lpzcr}, Zero Crossing Peak Amplitude (ZCPA) \cite {zcpa} and Pitch Synchronous Zero Crossing Peak Amplitude (PS-ZCPA) \cite {pszcpa}.
\item Amplitude: features are extracted from amplitude. An example is the Amplitude Descriptor (AD) that has been introduced for animal sounds discrimination \cite{ad}.
\item Power: it represents the mean square of the input raw signal such as, Short Time Energy (STE) \cite {ste} and volume \cite{volume}.
\end{itemize}

\subsection{Physical frequency features}
 The physical audio features are based on mathematical and statistical formulations such as, Fourier and Wavelet transforms. The physical frequency features are structured as follows: 
\begin{itemize}
\item Autoregression features: we can find features such as, Linear Predictive Coding (LPC) \cite {lpc} and Line Spectral Frequencies (LSF) \cite{lsf}.
\item Adaptive time-frequency decomposition features: they include features using time-frequency representations based on wavelet transformation. The advantage of the wavelet is the ability to provide  variable frequency resolutions within time \cite{mallat2008wavelet}.

\item Short time Fourier transform (STFT) features: these features calculated based on the STFT can capture properties of spectral envelope and phase information, such as subband energy ratio \cite{subband}, spectral flux \cite{sf}, spectral slope \cite{slope}, and spectral peaks \cite{peaks}.  
\end{itemize}

\subsubsection{Perceptual frequency features}

Contrary to physical features, the perceptual ones try to include the semantic in the feature extraction based on the human auditory system. The perceptual features are organized below:

\begin{itemize}
\item Brightness: brings information about the dominant frequency of the signal such as, spectral centroid \cite {spectralcentroid} and sharpness \cite {sharpness}.
\item Tonality: it is the characteristic of the sound that distinguish noise in tonal sounds including spectral dispersion \cite{dispersion} and spectral flatness \cite{flatness}.
\item Loudness: it includes integral loudness \cite {integralloudness}.
\item Pitch: several features have been introduced in this subgroup such as, pitch histogram \cite {pitchhistogram} and psychoacoustic pitch \cite{psychoacousticpitch}.
\item Chroma: the sensation of pitch is based on, tone height and chroma. The range of chroma is divided into 12 pitch classes such as the Pitch Class Profile (PCP) \cite{chroma}. 
\item Harmonicity: it represents the Power Spectral Density (PSD) at integer multiples of the fundamental frequency \cite{harmonicity}. 
\end{itemize}

\subsection{Cepstral features}
Cepstral features have been widely used in speech analysis. They aim to capture the timbral and pitch characteristics. We can find three main subgroups:

\begin{itemize}
\item Perceptual filter bank based features: they represent the Fourier transform of logarithm of the magnitude spectrum. A representative of these features is the widely used Mel-Frequency Cepstral Coefficients (MFCCs) and its extensions such as Relative Autocorrelation Sequence MFCC (RAS-MFCC) and CHNRAS-MFCC \cite{extensionmfcc}.
\item Advanced auditory model based features: these features try to model the physiological human hearing process. An example is noise robust audio features \cite{nraf}.
\item Autoregression based features: the features are calculated based on linear predictive analysis such as, Perceptual Linear Prediction (PLP) \cite{plp}, Relative Spectral Perceptual Linear Prediction (RASTA-PLP) \cite {rastaplp} and Linear Prediction Cepstrum Coefficients (LPCC) \cite{lpcc}.
\end{itemize}

\subsection {Modulation Frequency Features}
These features attempt to capture rhythm information. They represent a timbre and energy change over time such as, beat spectrum \cite{bandspectrum} and pulse metric \cite {sf}. \\


Table \ref {tab:features} summarizes different features along with their category and potential applications. The use of these presented features is not restricted to the reported applications. Extensions to other audio recognition tasks have been explored in the literature in order to evaluate their efficiency and genericity ability. The principal remark in this context is the fact that features designed for music were only successfully applied to music based application, in contrast to the speech and speaker recognition features which have already shown good performances for auditory scene recognition \cite{rakotomamonjy2015histogram}. This is due to the ability of speech-based features to capture intrinsic characteristics present in the audio scenes.


\begin{table*}
  \centering
  \caption {Overview of audio features and their applications. SP: Speech Recognition, SR: Speaker Recognition, CASR: Computational Auditory Scene Recognition, MA: Music Analysis, AR: Animal Sound Recognition.}
  \begin{tabular}{lll}
    \multicolumn{3}{c}{}  \\
        \toprule
    Type & Examples & Application   \\
    \midrule
   \textbf{1.~Temporal features} \\
    \tabitem Zero crossings & ZCR, LP-ZCR, ZCA, PS-ZCA & SP, SR, CASR \\
    \tabitem Amplitude & AD & AR  \\
     \tabitem Power & STE, Volume  & CASR    \\
     \textbf{2.~Physical frequency features} \\
      \tabitem Autoregression & LPC, LSF &  SP, SR, CASR  \\
       \tabitem Adaptive time-frequency decomposition & DWCH, ATFT & MA  \\
        \tabitem Short time Fourier transform & Spectral flux/slope/peaks & MA \\
        \textbf{3.~Perceptual frequency features} \\
          \tabitem Brightness & Spectral Centroid Sharpness & MA \\
        \tabitem Tonality & Spectral flatness/dispersion & MA \\
         \tabitem Loudness & Integral loudness & CASR \\
        \tabitem Pitch & Pitch histogram/psychoacoustic & MA  \\
        \tabitem Chroma crossings & PCP & MA \\
        \tabitem Harmonicity & PSD & MA \\
        \textbf{4.~Cesptral features} \\
         \tabitem Perceptual filter bank & MFCC, RAS-MFCC, CHNRAS-MFCC & SP, SR, CASR  \\
        \tabitem Advanced auditory model based & Noise robust & SP, SR\\
        \tabitem Autoregression based & PLP, RASTA-PLP, LPCC & SP, SR, CASR  \\
        \textbf{5.~Modulation frequency features} \\
        \tabitem Rythm & Beat spectrum, Pulse metric & MA  \\
    \bottomrule
  \end{tabular} \\ 
  \label{tab:features}
\end{table*}

\section{Human Behavior Analysis and Recognition Features Extraction}
\label{5}

Recognizing complex human behaviors and activities from video recorded data helps to develop intelligent video monitoring systems. However human behavior analysis and recognition represents one of the most challenging problems in the domain of computer vision due to the view angle variations, occlusions and the randomness of the activities. In visual perception based systems, the features try to capture characteristics that describe the human object segmented out from the raw video sequence such as, shape, silhouette, colors, poses, and body motions \cite{rida2018novel}. 

\begin{figure}[!ht]
\centering
\includegraphics [width= 9 cm] {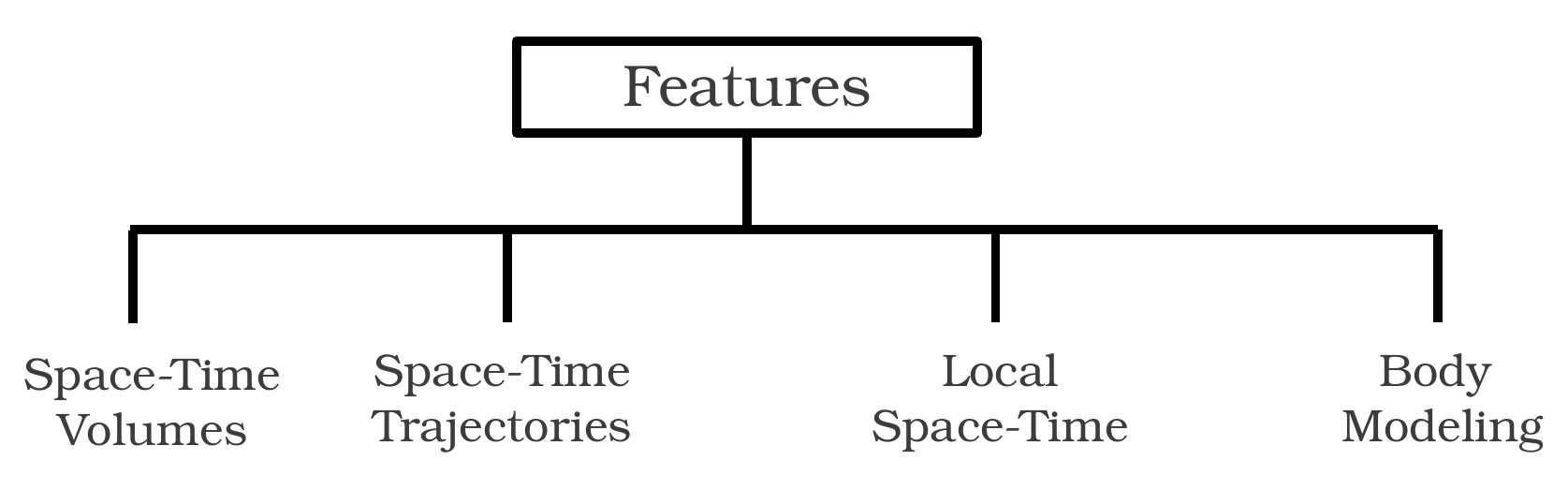}
\caption{Taxanomy of human behavior analysis and recognition features.}
\label{fig:taxanomybehavior}
\end {figure}

We introduce a taxonomy which divides these features into four main groups: space-time volumes, space-time trajectories, space-time local and body model as is shown in Figure \ref {fig:taxanomybehavior}. The next subsections describe those features.

\subsection {Space-Time Volumes}
\begin{figure}[!ht]
\centering
\includegraphics [width= 9 cm] {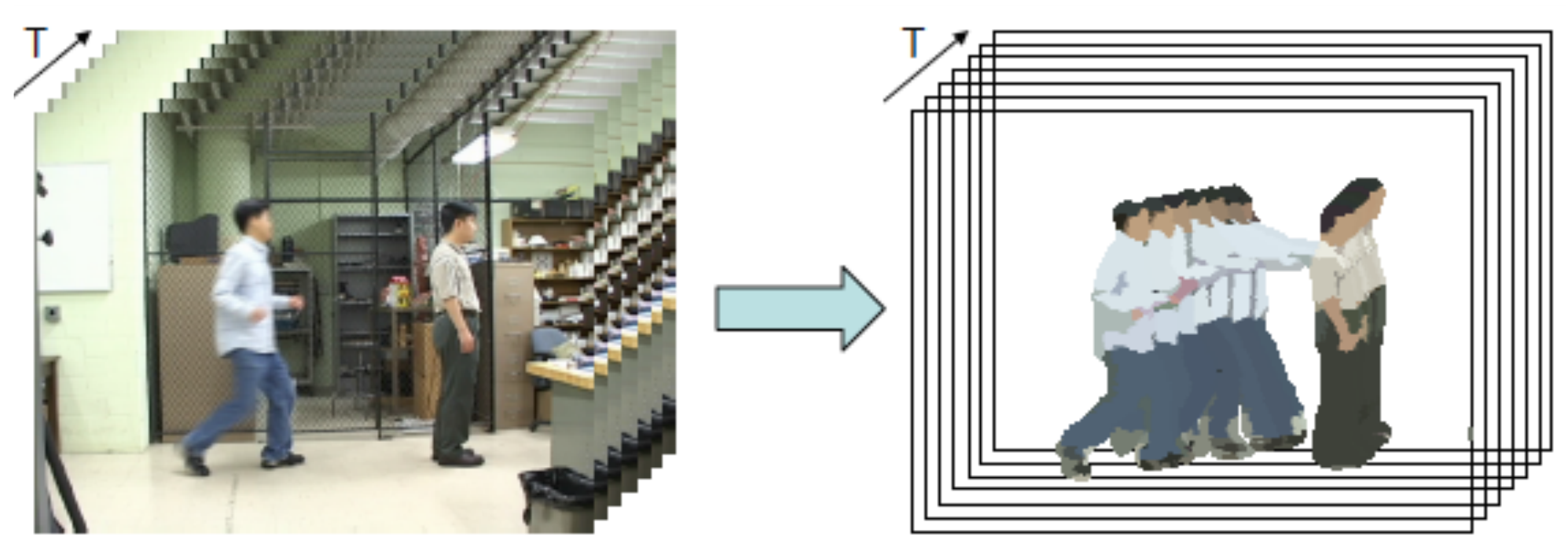}
\caption{An example of the space-time volumes construction \cite{aggarwal2011}.}
\label{fig:cube}
\end {figure}

Space-time volumes are constructed by stacking 2-D (XY) image frames along the time axis (T) as a 3D (XYT) cube as shown in Figure \ref{fig:cube}. The space-time volumes are able to capture both spatial and temporal information of the recorded object. Mainly the images are stacked after a segmentation step which aims to track the shape changes of the person in question \cite {bobick2001}. Based on the training video data, a space-time volume is constructed for different activities and persons \cite {shechtman2005,ke2007}.

Mainly, the space time volume features provide an efficient way to capture and combine both spatial and temporal information; however this requires a good preprocessing step of silhouette segmentations. Furthermore, viewpoint and occlusion are factors that drastically affect the performances.

 \subsection {Space-time trajectories}
\begin{figure}[!ht]
\centering
\includegraphics [width= 9 cm] {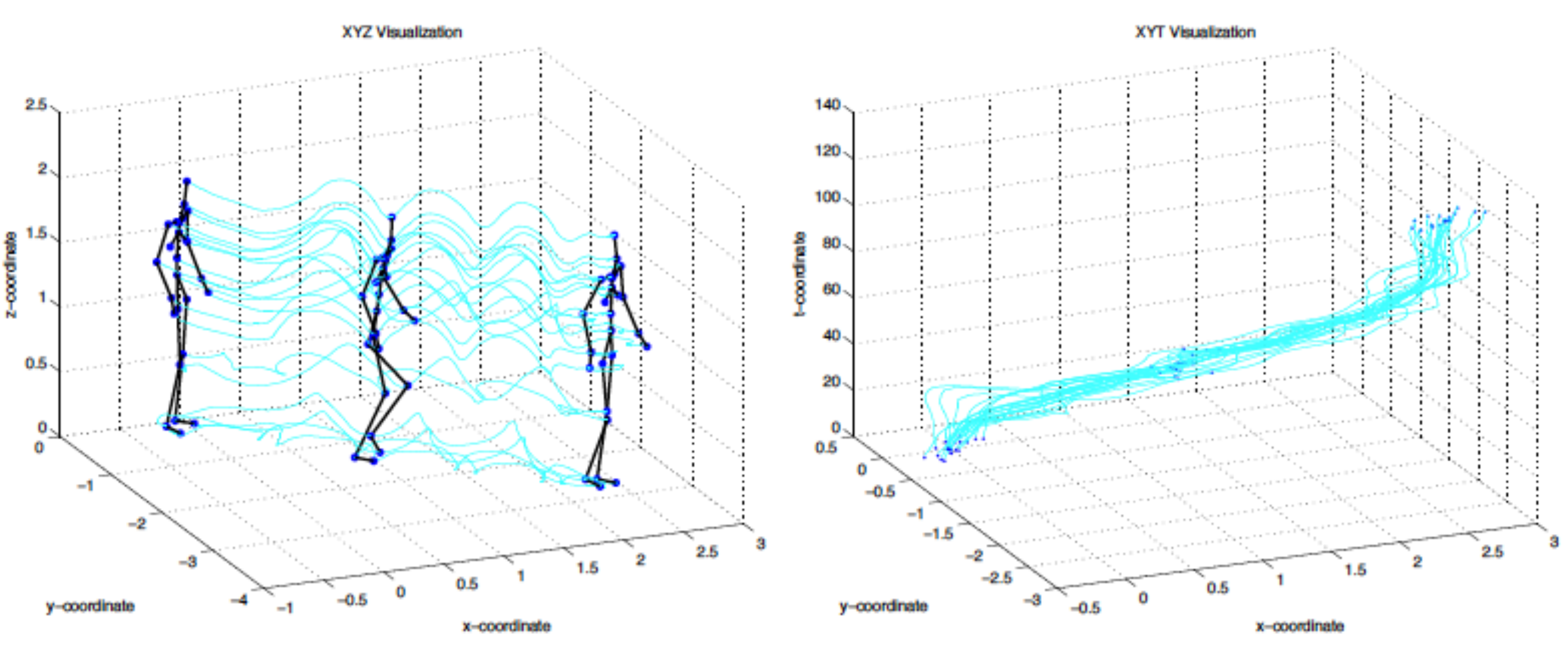}
\caption{An example of trajectories in XYZ and XYT spaces \cite{sheikh2005}.}
\label{fig:trajectory}
\end {figure}

These features seek to capture space-time trajectories by capturing the human joint positions as a set of 2-dimensional (XY) or 3-dimensional (XYZ) points. The trajectories are tracked over time which results 3-D XYT or 4-D XYZT representations as shown in Figure \ref {fig:trajectory}. Several works have used these features \cite{niyogi1994, rao2001, yilma2005}.

\subsection {Local Space-Time}
\begin{figure}[!ht]
\centering
\includegraphics [width= 5 cm] {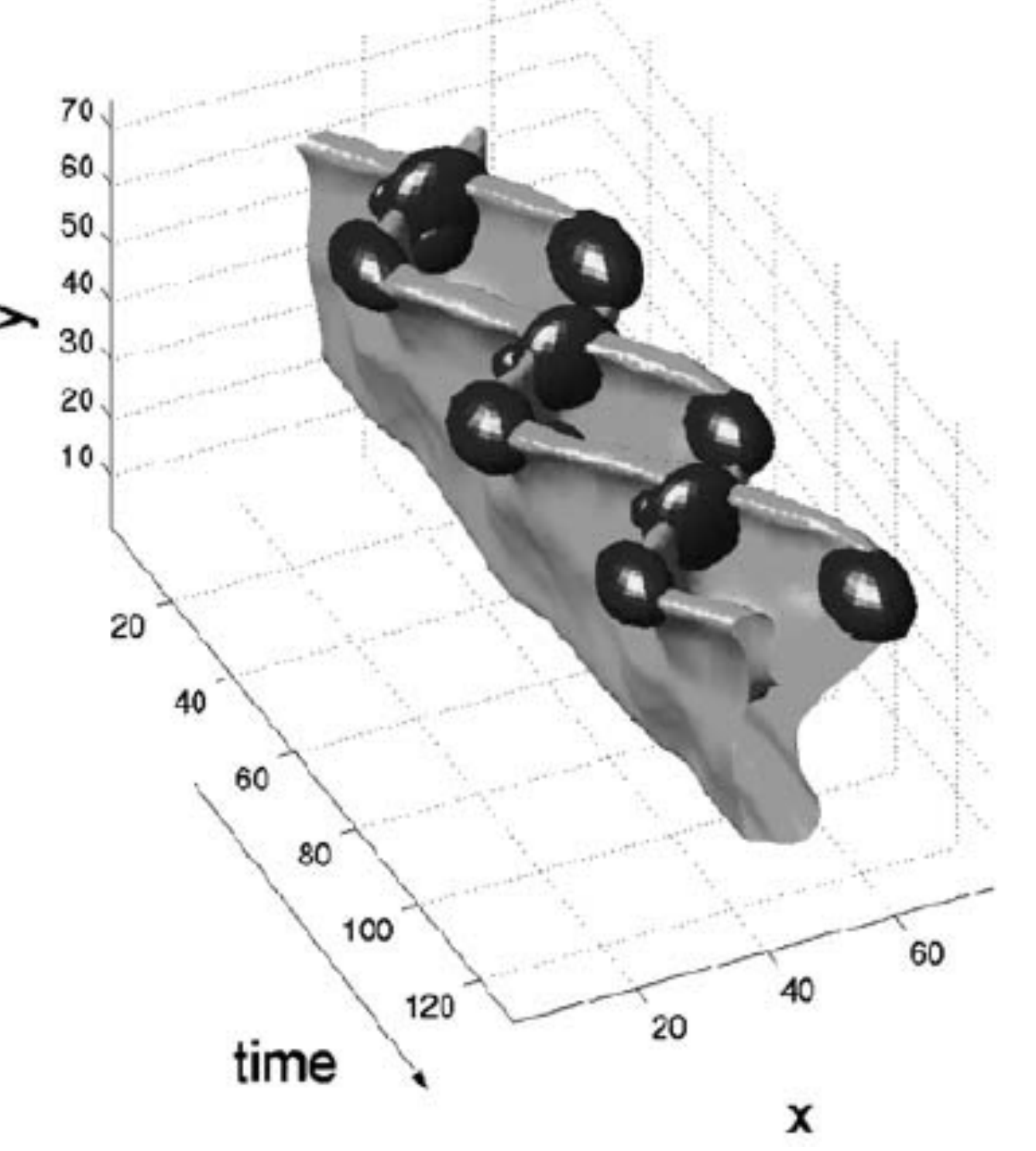}
\caption{An example of 3-D volumes (XYT) used to extract local features \cite{laptev2003}.}
\label{fig:laptev}
\end {figure}

3-D space-time volumes are considered as solid objects. This gives the ability to extract some appropriate local characteristics to distinguish between them. Several approaches are used to extract the local features: in \cite {chomat1999,zelnik2001,blank2005}, the local features are extracted from each video frame, the resulting features are concatenated over time to describe the human motion. In the other hand, some approaches extract local features directly from the 3-D volumes as is shown in Figure \ref {fig:laptev} \cite{laptev2003,dollar2005,niebles2008}. The local features are extracted using interest point detectors and descriptors such as, Harris operator, Laplacian of Gaussian (LoG), Scale-Invariant Feature Transform (SIFT) and Histogram of Oriented Gradients (HOG). 

\subsection {Body Modeling}

\begin{figure}[!ht]
\centering
\includegraphics [width= 5.5 cm] {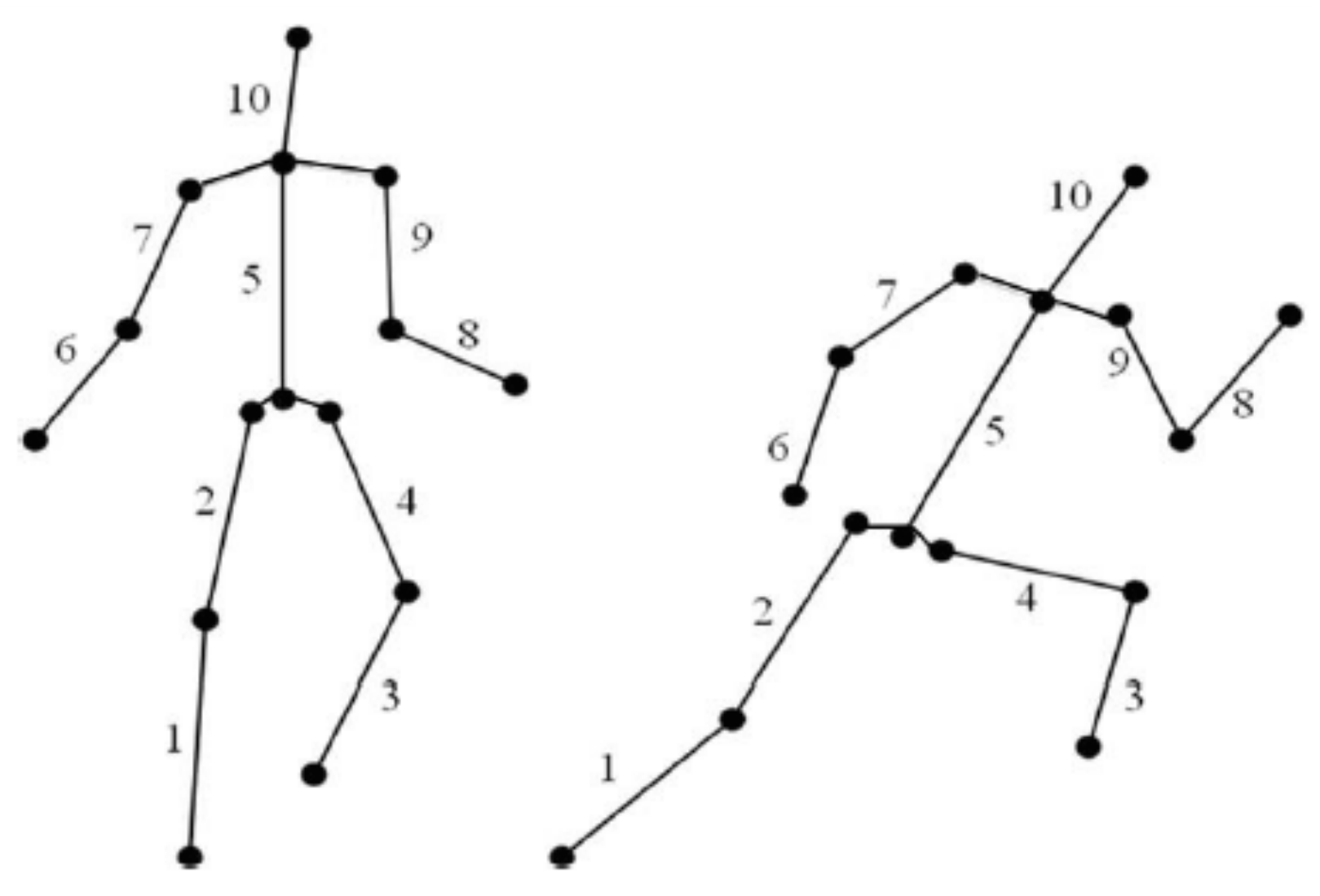}
\caption{An example of human body skeleton model \cite{sedai2009context}.}
\label{fig:skeleton}
\end {figure}

A human body model is developed to capture the 3D  geometric and kinematic structure of human body (see Figure \ref{fig:skeleton}). The model is supposed to extract information such as degrees of joint angles, length, width etc \cite{rida2017improve}. There have been several works using such features \cite{turaga2008,rogez2007}.

The previously introduced features for human behavior analysis and recognition try to capture the intrinsic characteristics of the moving subject and track their evolution over time \cite{rida2014improved}. In the space-time volumes, whole body silhouette is taken in consideration, it is simple to implement. However in outdoor conditions the subjects suffer from different intra-class variations caused by different conditions such as occlusion which make the segmentation step very complicated \cite{rida2016robust}. The performance of space-time volume features is affected by the quality of segmentation and can lead to very low performances in case of poor segmentation. Features based on space-time trajectories follow the same principle of the latter ones, however instead of taking the whole silhouette, some key points are retained to construct the moving body trajectories. The performance depends on the choice and amount of the trajectories. 

Motivated by the impact of segmentation on the performance of previous features, local space-time features have been introduced; they are extracted as local descriptors and are further concatenated to construct a feature vector. Following the same idea,  features capturing geometric and kinematic structure of the human body have been suggested. This type of features showed good performances however modeling the body is not a trivial task.


Once the features are extracted, finding a  suitable feature representation space is of extreme importance to achieve good classification performances. The main techniques of feature representations have been studied in \cite{rida2018comprehensive}.

\section{Conclusion}
\label{6}

 The recorded data represents a huge and important resource of information and knowledge which could be exploited in different real life applications including security, education, healthcare etc. Temporal signals have known a particular interest thanks to their ability to capture the intrinsic characteristics evolution over time. Indeed, temporal signals can be from different nature such as, gait, auditory scene or even a piece of music. In this paper, we have presented a general overview of the different extracted features by proposing simple taxonomies. This intended to help researchers to identify the most prominent features for different temporal signal recognition applications.


\end{document}